\documentclass{elsart}
\usepackage{natbib}

\newcommand{\um}{\,$\mu$m}
\def\lea{\mathrel{\raise .4ex\hbox{\rlap{$<$}\lower 1.2ex\hbox{$\sim$}}}}
\def\gea{\mathrel{\raise .4ex\hbox{\rlap{$>$}\lower 1.2ex\hbox{$\sim$}}}}

\begin{document}
\runauthor{Barthel \& van Bemmel}
\begin{frontmatter}
\title{Radio galaxies: unification and dust properties}
\author[]{Peter Barthel,}
\author[]{Ilse van Bemmel}

\address{Kapteyn Astronomical Institute, P.O.Box 800, NL--9700 AV Groningen}

\begin{abstract}

The 2002 status of unification models for extragalactic radio sources is
examined, with particular emphasis on the dust properties of
these objects. 

\end{abstract}

\begin{keyword}
radio galaxies; unification; dust; torus
\end{keyword}
\end{frontmatter}

\vspace{-10mm}

\section{Introduction}
\vspace{-6mm}
Unification of active galaxies combines the mechanisms of relativistic
beaming in radio jets (when present -- the radio-loud objects) with
anisotropy due to dust shadowing or non-spherical optically thick
emission; key parameter is the aspect angle. These geometric effects,
in interplay with source evolution -- youth, adulthood, seniority, as
well as duty cycle -- provide the full framework in which we seek to
explain the active galaxy populations. 

In case of Seyfert galaxies, an optically thick torus was invoked in the
mid 1980's to explain the apparent differences (strength of continuum
and emission line radiation) between Seyfert's of Types 1 and 2.  As for
the radio-loud population of radio galaxies and quasars such tori were
postulated in combination with relativistic effects in their radio
jets.  The model whereby quasars and BLLac objects are favourably
oriented radio galaxies has drawn considerable interest: the unification
church still has many members.  The review articles by \cite{ant93} and
\cite{up95} provide excellent accounts of these models.  Increased
attention has been paid in the last decade to the subject of life-time
evolution of active galaxies and AGN: some of that work will be dealt
with here, while \cite{bir02} provide an excellent full account. 
For a nice up-to-date summary of the general aspects of unification
studies we refer to the Proceedings of the Elba workshop "Issues in
Unification of AGN" \citep{elba02}. 


In this short review we will discuss new evidence from the past decade
in favour or against unification of active galaxies.  Furthermore, we
will examine the infrared properties of active galaxies, in order to
gain insight in the physical behaviour of the dusty toroid. 

\vspace{-5mm}
\section{Status of radio-loud unification models}
\vspace{-6mm}
Aspect being the key parameter in geometric unification, how do we know
the aspect angle to or the inclination of an active galactic nucleus?
Here radio-loudness helps because the fractional radio core strength
(radio core strength normalized with total radio or extended radio
luminosity) -- obviously taken from kpc-resolution radio
images -- provides a reasonably good orientation indicator. 
As pointed out by \cite{wb95}, the optical core luminosity normalized
with the total radio luminosity provides an additional, improved 
orientation indicator. In
addition, radio-loud objects provide us with a radio (jet) axis: radio
images permit determination of the projected source axis so only the
inclination angle w.r.t. this jet axis is unknown.  While radio-quiet
objects occasionally display optical cone emission yielding the optical
axis, that axis is generally less well constrained. 

As such, unification models for radio-loud objects are further developed
and encompass more seemingly different classes.  Within the framework of
this radio-loud meeting we will concentrate on the radio-loud
populations, but mention results for radio-quiet objects where relevant.

\vspace{-5mm}
\subsection{FR\,I unification}
\vspace{-6mm}
The model whereby Fanaroff \& Riley Class~I radio galaxies at small
inclination manifest themselves as BL\,Lacertae objects has gained
considerable support from HST observations of the former. 
Ultraviolet/optical cores in FR\,I host galaxies were found to correlate
with their radio cores arguing for a common, beamed synchrotron origin
\citep{cap02}.  No evidence for slower milli-arcsec scale jets in
comparison to FR\,II radio galaxies was found: FR\,I jets -- or their
relativistic spines -- must slow down from the parsec to the kiloparsec
scale \citep[e.g.,][]{giov01}.  Dust disks are often observed in FR\,I
hosts galaxies: their optical depth is much lower as compared to the
opaque circumnuclear tori postulated in FR\,II radio galaxies
\citep{chiab02}.  Added to the apparent absence of broad line emission
in FR\,I radio galaxies \cite[at least: luminous BLR -- 
see e.g.,][]{corb00}, this may mark an important distinction between the
Fanaroff \& Riley classes, but the distinction is blurred 
\citep[e.g.,][]{br01}. Further evidence is provided by the
far-infrared SEDs: FR\,I's are generally less far-infrared-bright than
FR\,II's at comparable radio luminosity \citep{heck94}, 
by a moderate factor of $\sim$4.  Note that
the broadband optical photometric host properties differ little or
nothing \citep{ledlow02} and that the masses 
of the central black holes
have no connection to radio luminosity whatsoever \citep[e.g.,][]{woo02}
-- it is most likely the accretion mechanism itself that will determine
the FR nature of the radio galaxy. 

Backyard FR\,I Centaurus~A was studied in quite some detail and found to
conform to the unification model \citep[e.g.,][]{capi00,chia01}.
\cite{why02} did however point out differences between archetypal
FR\,I's Centaurus~A and Virgo~A (M\,87). 

\vspace{-5mm}
\subsection{FR\,II unification}
\vspace{-6mm}
\cite{pdb93} found substantial supportive evidence for the favourable
orientation of all radio-loud quasars; most of this evidence has become
stronger.  This is however {\it not} equivalent with the picture whereby
all FR\,II radio galaxies contain a QSO hidden from direct view.  In
fact, the evidence for the existence of a population of FR\,II
pure-radio-galaxies -- without a big blue bump -- has grown.  Such a 
population of optically "dull"
FR\,II radio galaxies, in which the nuclear accretion activity is currently
switched off or at a low level, together with the astrophysically 
attractive model
of a receding torus (e.g., Simpson in these Proceedings) may well
account for the reported number density and linear size incompatibilities
between FR\,II radio galaxies and quasars. 

Investigations into possible orientation invariants, \`a la \cite{hes96}
and \cite{bak97} continue to be important.  Relatively weak [OIII]
emission may be due to obscuration but also to a weak ionizing spectrum
\citep[e.g.,][]{tadh98}.  With regard to the invariance issue, the mid-
and far-infrared emission remains controversial.  It is still not clear
to what extent this emission is suitable to test and/or constrain
unification models \citep[e.g.,][]{ivb00}; see Sect.~5.  The
multi-component, partly anisotropic nature of the far-infrared radiation
nevertheless does not provide major inconsistencies with unification
models; it is however likely that a cool dust component, related to host
star-formation activity plays a significant additional role. 

Beautiful radiation cones have been reported, in objects varying from
the nearest Seyfert galaxies to high redshift radio galaxies.  These
provide clear evidence for anisotropic nuclear radiation fields, with an
added component of jet driven star-formation (of difficult to determine
magnitude).  The strength of the latter is -- not surprisingly --
related to the size of the radio source \citep[e.g.,][]{best00}. 
However, in the powerful radio galaxy Cygnus~A there is solid
evidence that the ionization cones seen in optical images are
generated by an outflow, driven by the radiation pressure of the
central quasar \citep{ivb_phd}.

As for the host galaxies, the HST studies carried out by the Edinburgh
group have yielded strong support for the FR\,II unification, and also the
K($z$) behaviour provided consistency -- see the contribution by McLure in
these Proceedings.  HST and ground-based studies of radio galaxy hosts
by de Vries and
collaborators provide in addition consistency with evolutionary models
for the growth of radio sources; see Sect.~3. 

Obscured AGN are required by the X-ray background: this was already
pointed out in the late 1980's and the evidence is still strong 
\citep[e.g.,][]{comas95}.  X-ray spectra will soon yield the cosmologically
evolving obscuration, including the contribution from highly obscured
AGN \citep[Fabian, in][]{elba02}.

Spectropolarimetry provides the tool to {\it proof} unification. Following
up on the beautiful early work by Miller, Antonucci c.s., both
the Caltech and the Hatfield group obtained these proofs, for an as yet
small number of objects.  Polarization and obscuration/reddening appear
to go hand-in-hand, cf.  the models \citep[e.g.,][]{young96,mhc99}. 

Observations of the archetypal FR\,II narrow-line radio galaxy
Cygnus~A were wonderfully revealing: its radiation cones were imaged
with HST \citep[e.g.,][]{jack98, tadh99} whereas the hidden BLR was
detected with spectropolarimetry \citep{ogle97}.  We stress however
that ultraluminous backyard radio source Cygnus~A harbours a QSO of 
only moderate strength!

\vspace{-5mm}
\section{Effects of source evolution}
\vspace{-6mm}
Considerable effort was spent to study the nature of compact radio
sources.  These objects, of the Gigahertz-Peaked Spectrum (GPS) and
Compact Steep-Spectrum (CSS) classes, are now thought to represent the
progenitors of the large classical doubles.  Whereas the latter display
radio structure of supergalactic ($\gea 100$kpc) dimensions, the former
are subgalactic -- typically a few tens of kpc for the CSS and a few
tens of pc for the GPS class.  See also the contribution by Snellen in
these Proceedings.  Noteworthy is the determination of the spectral ages
of several CSS objects \citep{murgia99} which appear in agreement with
their postulated youth.  Cold H\,I gas has been detected in several CSS
radio galaxies (e.g., the contribution by Vermeulen in these
Proceedings); the CSS quasar class displays pronounced associated CIV
absorption \citep{baker02}.  Combined with the fact that some CSS radio
galaxies and quasars radiate unusually strong far-infrared emission this
has been postulated to imply a young evolutionary stage with strong
star-formation \citep{baker02}.  Well-known CSS quasar 3C\,48 provides
an excellent example \citep{can00}.  It is likely that the radio jets,
trying to find their way through the circumnuclear ISM play an important
role \citep[e.g.,][]{odea02}.  True (proper motion) expansion velocities,
of order 100 km/sec, for compact radio galaxies have been measured,
arguing for their very young ages \citep[e.g.,][]{iza99} and a nice 
case of a reborn GPS in a large double
lobed radio galaxy was recently reported \citep{mari02}.  Radio source
number density data require an increasing expansion speed and/or
decreasing radio luminosity with age -- to tie this down, the
evolutionary models for extragalactic radio sources are currently being
investigated with larger samples.  Overall, the global host galaxy
properties of compact and large scale radio galaxies do not show
inconsistencies with the evolutionary models \citep{wimdev00}. 

Support for the occurrence of recurrent nuclear activity (duty cycle ?)
is slowly accumulating \citep[e.g.,][]{arno00}.  Given the general
occurrence of massive black holes in luminous galaxies 
\citep[e.g.,][]{mago98}, such repetitive activity triggered by 
repetitive fueling
is not unexpected.  A multi-wavelength approach including deep optical
imaging seems at order. 

Broad-line radio galaxies make up an important subset of the radio
galaxy population.  Their nature is most likely composite: the class
encompasses low-luminosity QSRs, possibly with different torus opening
angles as compared to high luminosity QSRs, as well as radio galaxies
with somewhat transparent, porous tori \citep{tadh98,jdt00}.
\cite{ivb01} point out that BLRGs are characterized by the absence of
star-formation: one explanation could be that some BLRG represent old,
dying radio galaxies.  The absence of optical synchrotron (jet)
components in some BLRGs \citep{chiab02} may be consistent with that view. 

\vspace{-5mm}
\section{Very dusty sources}
\vspace{-6mm}
Intriguing recent development is the detection in X-rays of
active nuclei in seemingly non-active galaxies, apart from
their dusty starburst nature: NGC4945 \citep[e.g.,][]{guai00}
and NGC6240 \citep{iwa98}. Such highly obscured AGN
are important in the possible evolutionary connection between
starburst galaxies and AGN and may be important contributors
to the X-ray background -- ongoing Chandra and XMM-Newton
investigations will undoubtedly shed light on these issues.
 
Noting that dust obscuration lies at the heart of unification
models, we proceed by reviewing the status of torus models.

\vspace{-5mm}
\section{Modelling the toroid in AGN}


\vspace{-5mm}
\subsection{Introduction}
\vspace{-6mm}
Key to the unification model is the obscuring torus, creating an angle
and wavelength dependent anisotropy in active galaxies.  The soft X-ray
and UV emission from the central engine is reprocessed to infrared
wavelengths by the dust in the torus.  In 1983 IRAS was the first
satellite to detect infrared emission from the relatively distant active
galaxies.  Soon after, several groups started to model this
reprocessing using radiative transfer codes.  Among the first to
present their models are \cite{pk92}, hereafter PK92.  A few years 
later they are followed by \cite{gra94}, hereafter GD94 and 
\cite{efst95}, hereafter ER95, and recently models
have appeared by \cite{nenk02} and \cite{ivb03}, hereafter BD03. The 
ISO satellite
provided a wealth of additional data, however, only for the Seyfert
galaxies these allow a good constraint of the models. For the radio-loud
AGN population only few objects have a well defined broad-band SED,
e.g. Cygnus~A.

\vspace{-5mm}
\subsection{The early models: PK92, GD94 and ER95}
\vspace{-6mm}
To ease the radiative transfer calculations, all groups assume 
azimuthal symmetry, but the actual torus geometries differ among 
the different groups. PK92 define a
so-called pill-box geometry, where the thickness of the torus is 
constant with radius, and the inner walls are perpendicular to the 
plane of the toroid. GD94 use a similar geometry, but with
the possibility of having a conical hole in the center, instead of
perpendicular walls. They also allow for different dust mixtures
to be present in the torus. 
ER95 use three different geometries: a conical disk, where
the thickness of the disk increases linearly with radius,
an anisotropic sphere, and a pillbox with a conical hole.
In all their models, the torus has a constant inner radius, causing
the central opening to be circular (the opposite of the GD94 central
cavity). All these 'early' models assume
Galactic dust properties. 

\vspace{-5mm}
\subsection{The silicate problem}
\vspace{-6mm}
All three groups predict significant 10\um\ silicate emission
in type~1 active galaxies, which is not observed. The more powerful 
radio-loud population still lacks proper spectra, but the absence of 10\um\ 
emission is well established by the ISO spectrometers for the Seyfert 
type AGN. GD94 postulate depletion of small grains by shocks in order 
to explain the lack of 10\um\ emission in type~1's. Only one of the ER95 
pill-box model does not predict 10\um\ emission.

This leads several
groups to the conclusion that AGNs do not contain standard Galactic
dust. The grain size distribution might differ significantly from the
Galactic distribution, in the sense that AGN toroid dust is dominated by larger
grains. This is first recognized by \cite{laor93}, and several groups
have brought forth explanations for this.  From UV spectra it becomes
clear that also the 2200\,\AA\ absorption is shallower than expected in
many AGN, consistent with the lack of small grains \citep{maio01_2}. 

Small grains are easily destroyed in the strong radiation field of an
active nucleus.  However, they should survive in the regions shielded by
the torus.  Several reasons have been presented in literature for the
lack of small grains in AGN: most recently a clumping theory is
presented by \cite{maio01_1}.  The grains are swept up by the radiation
pressure, and clump together to form larger grains. 

\vspace{-5mm}
\subsection{The width problem}
\vspace{-6mm}
The early models encounter a second problem in explaining the full
width of the broad-band SED observed in all types of active galaxies.
Although the results from the models are generally broader than a single 
grey body, due to the temperature gradient in the dust, most of them 
are not yet broad enough. PK92 recognize this problem, and both GD94
and ER95 have tried to improve, but did not succeed.

\vspace{-5mm}
\subsection{A second component?}
\vspace{-6mm}
To deal with these issues, recent models have appeared in which the 
dust is no longer smooth, but clumpy \citep{nenk02}, or 
has a non-Galactic grain size distribution (BD03). 
Both sets of models can produce
the full observed width of the infrared SED of active galaxies, and
both manage to circumvent significant 10\um\ emission in models
at small inclinations. 

However, observational evidence is mounting that not all the infrared
emission in active galaxies arises from the compact torus 
\citep[see e.g.,][]{spi02,prie01}. There must
be a significant contribution from a second dust component, which seems
to be related to star-formation in the host galaxy, i.e. at scales much
larger than the radius at which the AGN can influence the dust. So far,
all papers have ignored this possibility, except BD03. Such a second
component
may well provide a solution to the width problem in the early models.
BD03 provide colour-colour diagrams for their models,
showing that the observed 25--60 and 60--100\um\ colours are not
well fitted by single component torus models, and demand an additional 
dust component.

This secondary dust component might also be responsible for the
lack of silicate emission observed in type~1 AGN; if the torus
does produce 10\um\ emission but this passes through a colder
layer of dust, the resulting spectrum shows no 10\um\ emission,
or even a shallow absorption. However, this would require a large
scale, spherical dust screen that covers all viewing angles to the
nuclear regions. In the case of NGC\,1068 there is no evidence of 
large scale silicate absorption: the nucleus seems to be the main 
source of the absorption. However, in this object the star-formation
provides only a marginal fraction of the total infrared luminosity
\citep{mar03}, whereas it is thought to contribute at 
least 50\% to the far-infrared emission in powerful 
radio sources.

For a proper understanding of dust in the nuclear regions of
AGN the relative contributions of large and small scale dust
need to be assessed. Decomposition will result in a better
understanding of the behaviour of dusty tori in active galaxies,
but also of the properties of the dust associated with star-formation
in the host galaxies. Although the current generation of models
is roughly consistent with observations, many important issues
need to be solved. For this purpose, deep SIRTF low resolution 
spectroscopy data will be most welcome.

\vspace{-5mm}
\section{Conclusions}
\vspace{-6mm}
Good progress has been made during the past decade regarding aspect
angle unification of radio-loud objects. The basic picture is correct,
but the distinction between various classes appears somewhat blurred.
The actual torus, big blue bump, and jet set-up, connected to
source power and lifetime evolution, is not yet understood.
The infrared might provide valuable clues, but we first need to
understand how the various components contribute to the overall SED.

{\it Acknowledgements}\\
We acknowledge expert reading by healthy Ski Antonucci noting in passing
that he fell ill just before the Workshop, leading the SOC to request
the present reviewers to take over Ski's review talk at 48h notice ..... 
The authors realize that this review is far from complete and does no
justice to many active workers in the field.  Apologies to those outside
the review cone .... 

\bibliographystyle{/ber1/users/bemmel/Tex/BibTex/aa}
\bibliography{/ber1/users/bemmel/Tex/BibTex/ivb}

\end{document}